\documentclass[12pt]{article}
\usepackage{epsf}

\begin{document}
\begin{titlepage}

\flushright{CLNS~01/1756\\
{\tt hep-ph/0110078}\\[0.2cm]
Revised Feb.\ 2002}

\vspace{1.2cm}
\begin{center}
\Large\bf\boldmath
Isospin Breaking in $B\to K^*\gamma$ Decays
\unboldmath
\end{center}

\vspace{0.5cm}
\begin{center}
Alexander L. Kagan\\[0.1cm]
{\sl Department of Physics, University of Cincinnati\\ 
Cincinnati, Ohio 45221, USA}\\[0.3cm]
and\\[0.3cm]
Matthias Neubert\\[0.1cm]
{\sl Newman Laboratory of Nuclear Studies, Cornell University\\
Ithaca, NY 14853, USA}
\end{center}

\vspace{0.8cm}
\begin{abstract}
\vspace{0.2cm}\noindent 
A calculation of the leading isospin-breaking contributions to the 
$B\to K^*\gamma$ decay amplitudes based on the QCD factorization 
approach is presented. They arise at order $\Lambda/m_b$ in the 
heavy-quark expansion and are due to annihilation contributions from 
4-quark operators, the chromo-magnetic dipole operator, and charm 
penguins. In the Standard Model the decay rate for 
$\bar B^0\to\bar K^{*0}\gamma$ is predicted to be about 10--20\% larger 
than that for $B^-\to K^{*-}\gamma$. Isospin-breaking effects are a
sensitive probe of the penguin sector of the effective weak Hamiltonian.
New Physics models in which the hierarchy of $B\to K^*\gamma$ decay 
rates is either flipped or greatly enhanced could be ruled out with more 
precise data.
\end{abstract}

\end{titlepage}

\section{Introduction}

The study of radiative decays based on the flavor-changing neutral 
current transition $b\to s\gamma$ is of crucial importance for testing 
the flavor sector of the Standard Model and probing for New Physics. 
Whereas the inclusive mode $B\to X_s\gamma$ can be analyzed using the 
operator product expansion, it is usually argued that exclusive decays 
such as $B\to K^*\gamma$ and $B\to\rho\,\gamma$ do not admit a clean 
theoretical analysis because of their sensitivity to hadronic physics. 
However, it has recently been shown that in the heavy-quark limit the 
decay amplitudes for these processes can be calculated in a 
model-independent way using a QCD factorization approach 
\cite{Bosch:2001gv,Beneke:2001at,Ali:2001ez}, which is similar to the 
scheme developed earlier for non-leptonic two-body decays of $B$ mesons 
\cite{BBNS}.

To leading order in $\Lambda/m_b$ (and neglecting isospin violation in
the $B\to K^*$ form factors) one finds that the amplitudes for the 
decays $\bar B^0\to\bar K^{*0}\gamma$ and $B^-\to K^{*-}\gamma$ 
coincide. Spectator-dependent effects enter at subleading order in the 
heavy-quark expansion. In this Letter the QCD factorization approach is 
used to estimate the leading isospin-breaking effects for the 
$B\to K^*\gamma$ decay amplitudes, the most important of which can be 
calculated in a model-independent way. 

Experimental measurements of exclusive $B\to K^*\gamma$ branching ratios 
have been reported by the CLEO, Belle and BaBar Collaborations, with the 
results (averaged over CP-conjugate modes):
\begin{eqnarray}
   10^5\,\mbox{Br}(\bar B^0\to\bar K^{*0}\gamma) 
   &=& \cases{ 4.55_{\,-\,0.68}^{\,+\,0.72}\pm 0.34 &
                ~\protect\cite{Coan:2000kh} \cr
               4.96\pm 0.67\pm 0.45 & ~\protect\cite{Ushiroda:2001sb} \cr
               4.23\pm 0.40\pm 0.22 & ~\protect\cite{Aubert:2001me} }
    \nonumber\\[-0.3cm]
    \\[-0.3cm]
   10^5\,\mbox{Br}(B^-\to K^{*-}\gamma) 
   &=& \cases{ 3.76_{\,-\,0.83}^{\,+\,0.89}\pm 0.28 &
                ~\protect\cite{Coan:2000kh} \cr
               3.89\pm 0.93\pm 0.41 & ~\protect\cite{Ushiroda:2001sb} \cr
               3.83\pm 0.62\pm 0.22 & ~\protect\cite{Aubert:2001me} }
    \nonumber
\end{eqnarray}
The average branching ratios for the two modes are 
$\mbox{Br}(\bar B^0\to\bar K^{*0}\gamma)=(4.44\pm 0.35)\cdot 10^{-5}$ 
and $\mbox{Br}(B^-\to K^{*-}\gamma)=(3.82\pm 0.47)\cdot 10^{-5}$. When 
corrected for the difference in the $B$-meson lifetimes, 
$\tau_{B^-}/\tau_{\bar B^0}=1.068\pm 0.016$ \cite{Blife}, these results 
imply
\begin{equation}
   \Delta_{0-}\equiv\frac{\Gamma(\bar B^0\to\bar K^{*0}\gamma)
                          -\Gamma(B^-\to K^{*-}\gamma)}
                         {\Gamma(\bar B^0\to\bar K^{*0}\gamma)
                          +\Gamma(B^-\to K^{*-}\gamma)}
   = 0.11\pm 0.07 \,.
\end{equation}
Although there is no significant deviation of this quantity from zero, 
the fact that all three experiments see a tendency for a larger neutral 
decay rate raises the question whether the Standard Model could account
for isospin-breaking effects of order 10\% in the decay amplitudes.

\section{Isospin-Breaking Contributions}

In the Standard Model the effective weak Hamiltonian for $b\to s\gamma$
transitions is
\begin{equation}
   {\cal H}_{\rm eff} = \frac{G_F}{\sqrt 2} \sum_{p=u,c} \lambda_p^{(s)}
   \bigg( C_1\,Q_1^p + C_2\,Q_2^p + \!\sum_{i=3,\dots,8}\! C_i\,Q_i 
   \bigg) ,
\end{equation}
where $\lambda_p^{(s)}=V_{ps}^* V_{pb}$ are products of CKM matrix
elements, $Q_{1,2}^p$ are the current--current operators arising from 
$W$ exchange, $Q_{3,\dots,6}$ are local 4-quark penguin operators, and 
$Q_7$ and $Q_8$ are the electro-magnetic and chromo-magnetic dipole 
operators. (We adopt the conventions of \cite{BBNS}; in particular, 
$C_1\approx 1$ is the largest coefficient.) The Wilson coefficients 
$C_i$ and the matrix elements of the renormalized operators $Q_i$ 
depend on the renormalization scale $\mu$. 

At leading power in $\Lambda/m_b$, and neglecting the tiny contribution 
proportional to $\lambda_u^{(s)}$, the $B\to K^*\gamma$ decay amplitude 
is given by
\begin{equation}\label{Alead}
   i{\cal A}_{\rm lead} = \frac{G_F}{\sqrt 2}\,\lambda_c^{(s)} a_7^c\,
   \langle\bar K^*(k,\eta)\gamma(q,\epsilon)|Q_7|\bar B\rangle \,,
\end{equation}
where the next-to-leading order (NLO) result for the coefficient 
$a_7^c=C_7+\dots$ is given in Eq.~(42) of \cite{Bosch:2001gv}. At this 
order contributions to the quantity $\Delta_{0-}$ arise only from 
isospin violation in the $B\to K^*$ form factors. (The phase-space 
difference between the two decays is an effect of order $(\Lambda/m_B)^2$
and contributes the negligible amount $-4\times 10^{-4}$ to 
$\Delta_{0-}$.) To first order we find
\begin{equation}
   \Delta_{0-}^{\rm soft}
   = 1 - \frac{T_1^{B\to K^{*-}}}{T_1^{B\to K^{*0}}} 
   \approx \frac{m_d-m_u}{m_s-m_d}\,\bigg( 1
   - \frac{T_1^{B\to\rho^0}}{T_1^{B\to K^{*0}}} \bigg)
   \approx 0.5\% \,,
\end{equation}
where $T_1^{B\to K^*}$ is a form factor in the decomposition of the
$B\to K^*$ matrix element of the tensor current 
$\bar s\sigma_{\mu\nu}(1+\gamma_5) b$ evaluated at zero momentum 
transfer. We have used the form-factor predictions of \cite{Ball:1998kk} 
and assumed that the ratio of isospin to $U$-spin violation for the 
transition form factors scales approximately as the corresponding 
ratio of light quark masses. Although this estimate is rather uncertain, 
we believe it indicates that the ``soft'' isospin-breaking effect is 
negligible and could not account for a value of $\Delta_{0-}$ as large 
as 10\%.

\begin{figure}
\epsfxsize=10.0cm
\centerline{\epsffile{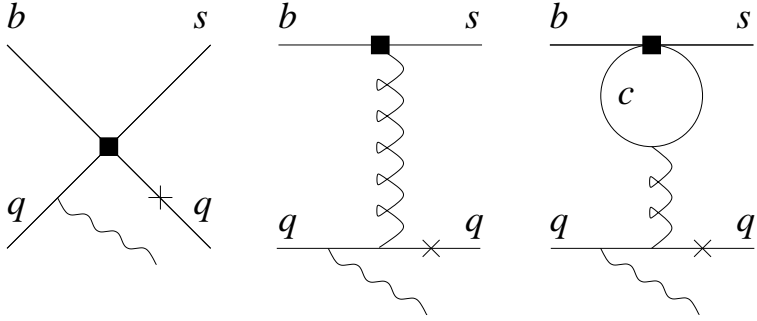}}
\vspace{0.2cm}
\centerline{\parbox{15cm}{\caption{\label{fig:graphs}
Spectator-dependent contributions from local 4-quark operators (left), 
the chromo-magnetic dipole operator (center), and the charm penguin 
(right). Crosses denote alternative photon attachments.}}}
\end{figure}

At subleading order in $\Lambda/m_b$ ``hard'' isospin-violating effects
appear in the form of spectator-dependent interactions. The leading 
contributions arise from the diagrams shown in Figure~\ref{fig:graphs}. 
For a first estimate of these effects we adopt a simplification of the 
usual NLO counting scheme, in which we neglect terms of order 
$\alpha_s\,C_{3,\dots,6}$ while retaining terms of order 
$\alpha_s\,C_{1,8}$. This is justified, because the penguin coefficients 
$C_{3,\dots,6}$ are numerically very small. Also, it is a safe 
approximation to neglect terms of order 
$\alpha_s\,\lambda_u^{(s)}/\lambda_c^{(s)}$ given that 
$|\lambda_u^{(s)}/\lambda_c^{(s)}|\sim 0.01$--0.02 is very small. It 
then suffices to evaluate the contributions of the 4-quark operators 
shown in the first diagram at tree level. The terms neglected in this 
simplified NLO approximation will be estimated later.

The isospin-breaking contributions to the decay amplitudes can be 
parameterized as ${\cal A}_q=b_q\,{\cal A}_{\rm lead}$, where $q$ is 
the flavor of the spectator antiquark in the $\bar B$ meson. To leading 
order in small quantities $\Delta_{0-}$ is then given by
\begin{equation}
   \Delta_{0-} = \mbox{Re}(b_d-b_u) \,.
\end{equation} 
The QCD factorization approach gives an expression for the coefficients 
$b_q$ in terms of convolutions of hard-scattering kernels with light-cone 
distribution amplitudes for the $K^*$ and $B$ mesons. When light quark 
masses are neglected, the leading and subleading projections for a 
transversely polarized vector meson with momentum $k$ and polarization 
$\eta$ are \cite{Ball:1998sk,Beneke:2001wa}
\begin{eqnarray}
   \langle\bar K^*(k,\eta)|\bar s(-z)_\alpha\dots q(z)_\beta|0\rangle 
   &=& \frac{f_{K^*}^\perp}{4}\,(\rlap/\eta^*\rlap/k)_{\beta\alpha}
    \int_0^1 dx\,e^{i\zeta k\cdot z}\,\phi_\perp(x) \nonumber\\
   &&\hspace{-5.3cm}
    \mbox{}+ \frac{f_{K^*} m_{K^*}}{4} \bigg[
    (\rlap/\eta^*)_{\beta\alpha}
    \int_0^1 dx\,e^{i\zeta k\cdot z}\,g_\perp^{(v)}(x) 
    - \frac12\,\epsilon_{\mu\nu\rho\sigma}\,
    \eta^{*\nu} k^\rho z^\sigma (\gamma^\mu\gamma_5)_{\beta\alpha}
    \int_0^1 dx\,e^{i\zeta k\cdot z}\,g_\perp^{(a)}(x) \nonumber\\
   &&\hspace{-3.0cm}
    \mbox{}+ \frac{\eta^*\!\cdot z}{k\cdot z}\,(\rlap/k)_{\beta\alpha}
    \int_0^1 dx\,e^{i\zeta k\cdot z} \left( \phi_\parallel(x)
    - g_\perp^{(v)}(x) \right) \! \bigg] \,,
\end{eqnarray}
where $z^2=0$, and the ellipses on the left-hand side indicate a string 
operator required to make the non-local matrix element gauge invariant. 
The variable $x$ is the longitudinal momentum fraction of the strange
quark, and $\zeta=(1-x)-x$. All four distribution functions are 
normalized to 1. The asymptotic form of the leading-twist amplitudes 
$\phi_{\perp,\parallel}(x)$ is $6x\bar x$ with $\bar x\equiv(1-x)$. In
the approximation where 3-particle distribution amplitudes of the kaon 
are neglected, the functions $\phi_\parallel$ and $g_\perp^{(v,a)}$ 
are related to each other by equations of motion 
\cite{Ball:1998sk,Beneke:2001wa}. Using these relations we find that 
(the prime denotes a derivative with respect to $x$)
\begin{equation}
   \frac{g_\perp^{(a)}(x)}{4\bar x} 
   + \frac{1}{\bar x} \int_0^x\!dy 
   \left( \phi_\parallel(y) - g_\perp^{(v)}(y) \right)
   = g_\perp^{(v)}(x) - \frac{g_\perp^{\prime\,(a)}(x)}{4} \,,
\end{equation}
which can be used to eliminate $\phi_\parallel$ from our results. This
means that we neglect contributions to the quantity $K_2$ arising from
$(g\bar q s)$ Fock states of the kaon, which are typically found to be 
suppressed with respect to two-particle contributions of the same
twist \cite{Ball:1998kk}. This approximation is justified, because 
numerically the effect of $K_2$ is about four times smaller than that 
of $K_1$. Including the 3-particle Fock states would, however, not 
invalidate factorization. 

The leading-twist projection onto the $B$ meson involves two 
distribution amplitudes $\Phi_{B1}(\xi)$ and $\Phi_{B2}(\xi)$ 
\cite{BBNS,Grozin:1997pq}, where $\xi=O(\Lambda/m_b)$ is the light-cone 
momentum fraction of the spectator quark projected onto the direction 
of the kaon. The first inverse moment of the function $\Phi_{B1}(\xi)$ 
defines a hadronic parameter $\lambda_B$ via
\begin{equation}
   \int_0^1 d\xi\,\frac{\Phi_{B1}(\xi)}{\xi} = \frac{m_B}{\lambda_B} \,.
\end{equation}
The function $\Phi_{B2}(\xi)$ does not enter our results. Note that 
3-particle Fock states of the $B$ meson only contribute to the quantity 
$K_1$ at order $\alpha_s(m_b)\,C_{5,6}(m_b)$ \cite{newpaper} and thus 
can be neglected in our approximation scheme.

The leading spectator-dependent contributions summarized by the 
coefficients $b_q$ can be written as
\begin{equation}\label{bqdef}
   b_q = \frac{12\pi^2 f_B\,Q_q}{m_b\,T_1^{B\to K^*} a_7^c}
   \left( \frac{f_{K^*}^\perp}{m_b}\,K_1
   + \frac{f_{K^*} m_{K^*}}{6\lambda_B m_B}\,K_2 \right) ,
\end{equation}
where $m_b$ denotes the running $b$-quark mass. The product 
$m_b\,T_1^{B\to K^*} a_7^c$ is scale independent at NLO. The 
heavy-quark scaling laws $f_B\sim m_b^{-1/2}$ and 
$T_1^{B\to K^*}\sim m_b^{-3/2}$ imply that $b_q$ scales like 
$\Lambda/m_b$, and hence the spectator-dependent corrections contribute 
at subleading order in the heavy-quark expansion. However, because of 
the large numerical factor $12\pi^2$ in the numerator the values of 
$b_q$ will turn out to be larger than anticipated in 
\cite{Bosch:2001gv,Beneke:2001at}.

The dimensionless coefficients $K_i$ are given by ($N=3$ and $C_F=4/3$ 
are color factors)
\begin{eqnarray}
   K_1 &=& - \left( C_6 + \frac{C_5}{N} \right) F_\perp \nonumber\\
   &&\mbox{}+ \frac{C_F}{N}\,\frac{\alpha_s}{4\pi}\,\left\{
    \left( \frac{m_b}{m_B} \right)^2 C_8\,X_\perp - C_1 \left[ \left(
    \frac43\ln\frac{m_b}{\mu} + \frac23 \right) F_\perp - G_\perp(s_c)
    \right] + r_1 \right\} , \\
   K_2 &=& \frac{\lambda_u^{(s)}}{\lambda_c^{(s)}}
    \left( C_1 + \frac{C_2}{N} \right) \delta_{qu}
    + \left( C_4 + \frac{C_3}{N} \right) 
    + \frac{C_F}{N}\,\frac{\alpha_s}{4\pi} \left\{ C_1
    \left( \frac43\ln\frac{m_b}{\mu} + \frac23 - H_\perp(s_c) \right) 
    + r_2 \right\} , \nonumber
\end{eqnarray}
where $r_1$ and $r_2$ are the residual NLO corrections neglected in our 
approximation scheme. The quantities 
\begin{eqnarray}
   F_\perp &=& \int_0^1 dx\,\frac{\phi_\perp(x)}{3\bar x} \,,
    \nonumber\\
   G_\perp(s_c) &=& \int_0^1 dx\,\frac{\phi_\perp(x)}{3\bar x}\,
    G(s_c,\bar x) \,, \nonumber\\
   H_\perp(s_c) &=& \int_0^1 dx\,\bigg( g_\perp^{(v)}(x) 
    - \frac{g_\perp^{\prime\,(a)}(x)}{4} \bigg)\,G(s_c,\bar x) \,,
    \nonumber\\
   X_\perp &=& \int_0^1 dx\,\phi_\perp(x)\,\frac{1+\bar x}{3\bar x^2}
\end{eqnarray}
with $s_c=(m_c/m_b)^2$ are convolution integrals of hard-scattering 
kernels with the meson distribution amplitudes, and
\begin{equation}
   G(s,\bar x) = -4\int_0^1 du\,u\bar u\,\ln(s-u\bar u\bar x-i\epsilon)
\end{equation}
is the penguin function. The terms in $K_2$ arising from the 
current--current operators $Q_{1,2}^u$ only contribute for $q=u$, as 
indicated by the symbol $\delta_{qu}$. Since no distinction between 
CP-conjugate modes is made in the experimental determination of 
$\Delta_{0-}$ we only need the real part of the ratio of CKM parameters,
given in terms of Wolfenstein parameters as $\mbox{Re}
(\lambda_u^{(s)}/\lambda_c^{(s)})=\lambda^2\bar\rho=0.011\pm 0.005$.

The first three convolution integrals above exist for any reasonable
choice of the distribution amplitudes. This shows that, to the order we 
are working, the QCD factorization approach holds at subleading power 
for the matrix elements of the 4-quark and current--current operators
(including penguin contractions). Factorization of these matrix elements
is indeed expected to hold to all orders in perturbation theory as a 
consequence of color transparency \cite{Bosch:2001gv}. This implies, in 
particular, that long-distance contributions to the first diagram in 
Figure~\ref{fig:graphs}, which have been analysed using QCD sum rules in 
\cite{Ali:1995uy,Khodjamirian:1995uc}, must be suppressed by at least 
two powers of $\Lambda/m_b$ and so do not contribute to the order we are 
working. On the other hand, if the function $\phi_\perp(x)$ vanishes 
only linearly at the endpoints (as indicated by its asymptotic 
behavior), then the convolution integral $X_\perp$ suffers from a 
logarithmic endpoint singularity as $x\to 1$, corresponding to the 
region where the light spectator in the $K^*$ meson is a soft quark. 
This indicates that at subleading power factorization breaks down for 
the matrix element of the chromo-magnetic dipole operator $Q_8$ (second
diagram in Figure~\ref{fig:graphs}). In the phenomenological analysis 
below we regulate this singularity by introducing a cutoff such that 
$x<1-\Lambda_h/m_B$, where $\Lambda_h=0.5$\,GeV is a typical hadronic 
scale. Since $X_\perp$ is dominated by soft physics, we assign a large 
uncertainty to this estimate.

The Wilson coefficients $C_{3,\dots,6}$ entering the expressions for 
$K_1$ and $K_2$ must be evaluated at NLO, while the remaining 
coefficients can be taken at leading order. Throughout, the two-loop 
expression for the running coupling $\alpha_s(\mu)$ is used. The 
explicitly scale-dependent terms in the expressions for the coefficients 
$K_i$ arise from the charm-penguin diagrams. They are necessary to 
cancel the scale and scheme dependence of the Wilson coefficients 
$C_{3,\dots,6}$. In a complete NLO calculation there would be additional 
logarithmic terms proportional to $\alpha_s\,C_{3,\dots,6}$, which can 
be deduced using the renormalization group (RG). This gives for the 
remainders
\begin{eqnarray}\label{r1r2}
   r_1 &=& \left[\, \frac83\,C_3 + \frac43\,n_f(C_4+C_6) - 8(N C_6+C_5)
    \right] F_\perp \ln\frac{\mu}{\mu_0} + \dots \,, \nonumber\\
   r_2 &=& \left[ -\frac{44}{3}\,C_3 - \frac43\,n_f(C_4+C_6) \right]
    \ln\frac{\mu}{\mu_0} + \dots \,, 
\end{eqnarray}
where $n_f=5$, and $\mu_0=O(m_b)$ is an arbitrary normalization point. 
The sensitivity to $\mu_0$ provides an estimate of the residual 
non-logarithmic NLO terms denoted by the ellipses, whose calculation is 
left for future work. After the addition of the $r_i$ pieces the 
quantity $K_2$ is RG invariant at NLO. In the case of $K_1$ a scale 
dependence remains, which cancels against the scale dependence of the 
tensor decay constant, $f_{K^*}^\perp(\mu)\sim
[\alpha_s(\mu)]^{C_F/\beta_0}$, and of the running $b$-quark mass. 

An important element of the phenomenological analysis are the 
convolution integrals. We adopt the shapes of the light-cone 
distribution amplitudes obtained (at second order in the Gegenbauer 
expansion) using QCD sum rules \cite{Ball:1998sk} and vary the amplitude 
parameters within their respective error ranges. This leads to 
$F_\perp=1.21\pm 0.06$, $G_\perp(s_c)=(2.82\pm 0.20)+(0.81\pm 0.23)i$ 
and $H_\perp(s_c)=(2.32\pm 0.16)+(0.50\pm 0.18)i$, where 
$m_c/m_b=0.26\pm 0.03$ has been used for the ratio of quark masses. 
Next, we find $X_\perp=(3.44\pm 0.47)\,X-(3.91\pm 1.08)$, where 
$X=\ln(m_B/\Lambda_h)\,(1+\varrho\,e^{i\varphi})$ parameterizes the 
logarithmically divergent integral $\int_0^1 dx/\bar x$. Following 
\cite{BBNS} we allow $\varrho\le 1$ and an arbitrary strong-interaction 
phase $\varphi$ to account for the theoretical uncertainty due to soft 
rescattering in higher orders. The above results for the convolution
integrals refer to a renormalization point of $\sqrt5$\,GeV. (The 
expressions for $r_i$ given earlier are valid if the convolution 
integrals are normalized at a fixed scale.)

Further input parameters are $m_b(m_b)=4.2$\,GeV, the decay constants 
$f_B=(200\pm 20)$\,MeV \cite{Penin:2001ux,Jamin:2001fw},
$f_{K^*}=(226\pm 28)$\,MeV and $f_{K^*}^\perp=(175\pm 9)$\,MeV (at 
$\mu=\sqrt5$\,GeV) \cite{Ball:1998sk}, and the parameter 
$\lambda_B=(350\pm 150)$\,MeV defined in terms of the first inverse 
moment of the $B$-meson distribution amplitude $\Phi_{B1}$ \cite{BBNS}. 
A dominant uncertainty in the prediction for $\Delta_{0-}$ comes from 
the tensor form factor $T_1^{B\to K^*}$, recent estimates of which 
range from $0.32_{\,-\,0.02}^{\,+\,0.04}$ \cite{DelDebbio:1998kr} to 
$0.38\pm 0.06$ \cite{Ball:1998kk}. On the other hand, a fit to the 
$B\to K^*\gamma$ branching fractions yields the lower value 
$0.27\pm 0.04$ \cite{Beneke:2001at}. To good approximation the result 
for $\Delta_{0-}$ is inversely proportional to the value of the form 
factor. Below we take $T_1^{B\to K^*}=0.3$ (at $\mu=m_b$) as a reference 
value. 

\begin{figure}
\epsfxsize=10.0cm
\centerline{\epsffile{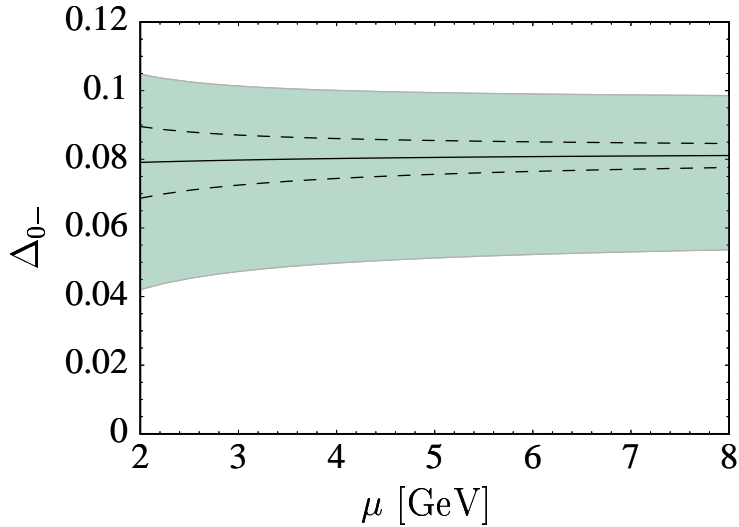}}
\vspace{0.2cm}
\centerline{\parbox{15cm}{\caption{\label{fig:result}
Prediction for the quantity $\Delta_{0-}$ as a function of the 
renormalization scale, assuming $T_1^{B\to K^*}=0.3$. The dark lines
refer to $\mu_0=m_b$ (solid), $m_b/2$ (upper dashed) and $2m_b$ (lower
dashed). The band shows the theoretical uncertainty.}}}
\end{figure}

From the diagrams in Figure~\ref{fig:graphs} it is seen that in all 
cases the operators are probed at momentum scales of order 
$\mu\sim m_b$. Hence, following common practice we vary the 
renormalization scale between $m_b/2$ and $2 m_b$. The result for 
$\Delta_{0-}$ is shown in Figure~\ref{fig:result}. The width of the band 
reflects the sensitivity to input parameter variations. The three curves 
correspond to different choices of the scale $\mu_0$ in the expressions 
(\ref{r1r2}) for the remainders $r_i$. The excellent stability under 
variation of both $\mu$ and $\mu_0$ shows that our approximation scheme 
captures the dominant terms at NLO. 

Combining all sources of uncertainty we obtain
\begin{equation}
   \Delta_{0-} = (8.0_{\,-\,3.2}^{\,+\,2.1})\% \times
   \frac{0.3}{T_1^{B\to K^*}} \,.
\end{equation}
The three largest contributions to the error from input parameter 
variations are due to $\lambda_B$ (${}_{\,-\,2.5}^{\,+\,1.0}\%$), the 
divergent integral $X_\perp$ ($\pm 1.2\%$), and the decay constant $f_B$ 
($\pm 0.8\%$). The perturbative uncertainty is about $\pm 1\%$. Our 
result is in good agreement with the current central experimental value 
of $\Delta_{0-}$ including its sign, which is predicted unambiguously. 
By far the most important source of isospin breaking is due to the 
4-quark penguin operator $Q_6$, whose contribution to $\Delta_{0-}$ is 
about 9\% (at $\mu=\mu_0=m_b$). The other terms are much smaller. In 
particular, the contribution of the chromo-magnetic dipole operator, for 
which factorization does not hold, is less than 1\% in magnitude and 
therefore numerically insignificant. Hence, the most important 
isospin-breaking contributions can be calculated using QCD factorization. 
It follows from our result that these effects mainly test the magnitude 
and sign of the ratio $C_6/C_7$ of penguin coefficients.

\section{New Physics}

Because of their relation to matrix elements of penguin operators, 
isospin-breaking effects in $B\to K^*\gamma$ decays are sensitive probes 
of physics beyond the Standard Model. For example, scenarios in which 
the sign of $\Delta_{0-}$ is flipped could be ruled out with more 
precise data. In addition, in certain extensions of the Standard Model 
there exist local 4-quark operators yielding an isospin-breaking 
contribution to the decay amplitudes at leading power in the heavy-quark 
expansion. Precise measurements of radiative decay rates would tightly 
constrain the corresponding Wilson coefficients. Here we confine 
ourselves to a brief illustration of the most interesting potential New 
Physics effects in $B\to K^*\gamma$ decays. A more detailed study will 
be presented elsewhere.

New Physics effects arising at some high energy scale manifest 
themselves at low energy through new contributions to the effective 
weak Hamiltonian. As a first, popular example in which the operator 
basis is not enlarged, consider the minimal supersymmetric Standard 
Model (MSSM) with minimal flavor violation, and with contributions to 
the $B\to X_s \gamma$ decay rates that are enhanced in the 
large-$\tan\beta$ limit taken into account beyond leading order 
\cite{degrassi,carena,demir}. In this scenario new contributions to 
$Q_{3,\dots,6}$ and $Q_8$ are too small to have a significant effect.
For low $\tan\beta$, $\mbox{Re}(C_7)$ is negative as in the Standard 
Model. However, for large $\tan\beta$ the coefficient $\mbox{Re}(C_7)$ 
can take on both positive or negative values. Positive values, which 
would flip the sign of $\Delta_{0-}$, become more probable as 
$\tan\beta$ increases (see, e.g., \cite{bozpak}). With more precise 
data, isospin breaking in $B\to K^*\gamma$ decays could rule out 
significant regions of the MSSM parameter space at large $\tan\beta$. 
We note that the sign of $\mbox{Re}(C_7)$ can also be flipped in 
supersymmetric models with non-minimal flavor violation, independently 
of $\tan\beta$, via gluino--down-squark loop graphs. 

In a more general scenario, New Physics effects can be parameterized in
terms of additional contributions to the coefficients of the dipole 
operators $Q_7$ and $Q_8$, contributions involving dipole operators 
$\widehat Q_7$ and $\widehat Q_8$ with opposite chirality than in the 
Standard Model, and a plethora of local 4-quark operators, the most 
general set of which can be taken as (a factor of 
$\frac{G_F}{\sqrt 2}\,\lambda_c^{(s)}$ is included here for convenience 
only)
\begin{equation}
   {\cal H}_{\rm eff,NP}^{\rm 4\mbox{-}quark} 
   = \frac{G_F}{\sqrt 2}\,\lambda_c^{(s)} \hspace{-0.2cm}
   \sum_{q=u,d,\dots}
   \,\sum_{\Gamma=\Gamma_1\otimes\Gamma_2}\!
   \left( c_\Gamma^q\,O_\Gamma^q
   + \widetilde c_\Gamma^q\,\widetilde O_\Gamma^q \right) ,
\end{equation}
where $O_\Gamma^q=\bar s\,\Gamma_1 b\,\bar q\,\Gamma_2 q$ and
$\widetilde O_\Gamma^q=\bar s_i\Gamma_1 b_j\,\bar q_j\Gamma_2 q_i$. The 
Dirac structure $\Gamma=\Gamma_1\otimes\Gamma_2$ can be any of the ten 
combinations $(V\mp A)\otimes(V\mp A)$, $(V\mp A)\otimes(V\pm A)$, 
$(S\pm P)\otimes(S\mp P)$, $(S\pm P)\otimes(S\pm P)$, and 
$T_{L,R}\otimes T_{L,R}$. Here, as usual, $S=1$, $P=\gamma_5$, 
$V=\gamma_\mu$, $A=\gamma_\mu\gamma_5$, and $T_{L,R}
=\sigma_{\mu\nu}(1\mp\gamma_5)$. As an example, note that such an 
extensive list of 4-quark operators can arise in general supersymmetry 
models \cite{Grossman:1999av,Borzumati:2000qt}.

In a generic New Physics model the Wilson coefficients $c_\Gamma^q$ and 
$\widetilde c_\Gamma^q$ need not be flavor independent 
\cite{Grossman:1999av}. Therefore, isospin-breaking effects can arise 
even if the photon in the first diagram in Figure~\ref{fig:graphs} is 
emitted from the $b$ or $s$-quark lines. Taking this possibility into 
account, we find that at leading order the contributions to the 
coefficients $K_i$ are
\begin{eqnarray}\label{KiNP}
   K_1^{\rm NP} &=& \frac{m_B}{6\lambda_B} \left( 
    \bar c_{T_R\otimes T_R}^q - \frac12\,\bar c_{(S+P)\otimes(S+P)}^q
    \right) - \bigg( F_\perp + \frac{Q_d}{Q_q}\,\bar F_\perp \bigg)
    \bar c_{(V-A)\otimes(V+A)}^q \nonumber\\
   &&\mbox{}+ \frac{C_F}{N}\,\frac{\alpha_s}{4\pi}\,X_\perp\,
    C_8^{\rm NP} \,, \nonumber\\
   K_2^{\rm NP} &=& \bar c_{(V-A)\otimes(V-A)}^q
    - \frac12\,\bar c_{(S-P)\otimes(S+P)}^q \,,
\end{eqnarray}
where $\bar c_\Gamma^q\equiv\widetilde c_\Gamma^q+c_\Gamma^q/N$, and we
have introduced the new convolution integral
\begin{equation}
   \bar F_\perp = \int_0^1 dx\,\frac{\phi_\perp(x)}{3x}
   = 0.84\pm 0.06 \,.
\end{equation}
The New Physics contribution from the chromo-magnetic dipole operator 
is included in (\ref{KiNP}) despite its $O(\alpha_s)$ suppression, 
because it could potentially be large in models with strongly enhanced 
$C_8$. In analogy with (\ref{bqdef}), we define a quantity
$\widehat b_q^{\rm NP}$ with corresponding coefficients 
$\widehat K_i^{\rm NP}$ given by equivalent expressions with all Wilson 
coefficients replaced by their opposite-chirality counterparts. Its 
contribution to the decay amplitude is 
$\widehat{\cal A}_q=\widehat b_q^{\rm NP}\,\widehat{\cal A}_{\rm lead}$, 
where $\widehat{\cal A}_{\rm lead}$ is defined in analogy with 
(\ref{Alead}) in terms of the matrix element of the opposite-chirality 
operator $\widehat Q_7$. Because this amplitude does not interfere with 
the leading Standard Model amplitude for $B\to K^*\gamma$ its effect is 
likely to be suppressed.

A remarkable fact is that there can exist local 4-quark operators 
yielding a leading (i.e., not power-suppressed) contribution to the 
decay amplitudes, as indicated by the factor $\sim m_B/\lambda_B$ in the 
first term in $K_1^{\rm NP}$ and $\widehat K_1^{\rm NP}$, which 
compensates the power suppression from the prefactor $f_{K^*}^\perp/m_b$  
in (\ref{bqdef}). The origin of these terms can readily be understood by 
noting that the tensor operators $\bar s\sigma_{\mu\nu}(1\pm\gamma_5)q\,
\bar q\sigma^{\mu\nu}(1\pm\gamma_5)b$ have a leading-twist projection 
onto the $K^*$ meson. When Fierz-transformed into the basis used above, 
such operators turn into tensor operators and operators with structure 
$(S\pm P)\otimes(S\pm P)$. The fact that such operators enter the 
amplitudes for exclusive radiative $B$ decays with large coefficients
means that future, precise measurements of radiative decay rates will 
provide tight constraints on the corresponding Wilson coefficients.

\section{Summary}

We have presented a model-independent analysis of the leading 
isospin-breaking contributions to the $B\to K^*\gamma$ decay amplitudes.
In the Standard Model these effects appear first at order $\Lambda/m_b$ 
in the heavy-quark expansion. They can be expressed in terms of 
convolutions of hard-scattering kernels and meson light-cone distribution
amplitudes of leading and subleading twist. We have evaluated these 
contributions at NLO in perturbation theory, neglecting however some 
numerically suppressed $O(\alpha_s)$ terms. With the exception of the 
matrix element of the chromo-magnetic dipole operator, whose 
contribution is numerically small, factorization of the leading 
isospin-breaking contributions is expected to hold to all orders in 
perturbation theory. 

To our knowledge, the analysis presented here provides the first example 
of a quantitative test of QCD factorization at the level of power 
corrections. As such, it lends credibility to the idea of factorization 
as a leading term in a well-behaved expansion in inverse powers of the 
$b$-quark mass. Our prediction for the magnitude and sign of 
isospin-breaking is in good agreement with the present central 
experimental value of this effect. If this agreement persists as the 
data become more precise, it will be possible to place novel constraints 
on flavor physics beyond the Standard Model.

\vspace{0.4cm}  
{\it Acknowledgments:\/} 
This work originated during the Workshop on Flavor Physics at the Aspen 
Center for Physics. We are grateful to M.~Beneke, G.~Buchalla and
A.~Petrov for useful discussions. A.K.\ thanks the SLAC Theory Group
for its hospitality and support, and the Department of Energy for 
support under Grant DE-FG02-84ER40153. M.N.\ is supported by the 
National Science Foundation under Grant PHY-0098631.

\end{document}